# INTEGRAL[1]: science highlights and future prospects

Christoph Winkler[1], Roland Diehl[2], Pietro Ubertini[3], Jörn Wilms[4]

*(1) European Space Agency, ESTEC, Keplerlaan 1, NL-2201 AZ Noordwijk, The Netherlands*
cwinkler@rssd.esa.int

*(2) Max-Planck Institut für Extraterrestrische Physik, Giessenbachstr, D-85748 Garching, Germany*
rod@mpe.mpg.de

*(3) INAF - IASF Roma, Via del Fosso del Cavaliere 100, I-00133 Roma, Italy*
pietro.ubertini@iasf-roma.inaf.it

*(4) Dr. Karl Remeis-Sternwarte Bamberg, Astronomisches Institut, Universität Erlangen-Nürnberg, Sternwartstraße 7, D-96049 Bamberg, Germany*
joern.wilms@sternwarte.uni-erlangen.de

*Keywords: high-energy astrophysics, black holes, neutron stars, nucleosynthesis, active galactic nuclei, gamma-ray bursts, space instrumentation*

Abstract

ESA's hard X-ray and soft gamma-ray observatory INTEGRAL is covering the 3 keV to 10 MeV energy band, with excellent sensitivity during long and uninterrupted observations of a large field of view (~100 $\Box^o$), with ms time resolution and keV energy resolution. It links the energy band of pointed soft X-ray missions such as XMM-Newton with that of high-energy gamma-ray space missions such as Fermi and ground based TeV observatories. Key results obtained so far include the first sky map in the light of the 511 keV annihilation emission, the discovery of a new class of high mass X-ray binaries and detection of polarization in cosmic high energy radiation. For the foreseeable future, INTEGRAL will remain the only observatory allowing the study of nucleosynthesis in our Galaxy, including the long overdue next nearby supernova, through high-resolution gamma-ray line spectroscopy. Science results to date and expected for the coming mission years span a wide range of high-energy astrophysics, including studies of the distribution of positrons in the Galaxy; reflection of gamma-rays off clouds in the interstellar medium near the Galactic Centre; studies of black holes and neutron stars particularly in high- mass systems; gamma-ray polarization measurements for X-ray binaries and gamma-ray bursts, and sensitive detection capabilities for obscured active galaxies with more than 1000 expected to be found until 2014. This paper summarizes scientific highlights obtained since INTEGRAL's launch in 2002, and outlines prospects for the INTEGRAL mission.

---

[1] INTEGRAL: *The International Gamma-Ray Astrophysics Laboratory*, an ESA project with instruments and science data centre funded by ESA member states (especially the PI countries: Denmark, France, Germany, Italy, Switzerland, Spain), Poland and with the participation of Russia and the USA.

# Introduction

Gamma-ray astronomy explores the most energetic phenomena that occur in nature and addresses some of the most fundamental problems in physics and astrophysics. It embraces a great variety of gamma-ray continuum and gamma-ray line processes: nuclear excitation, radioactivity, positron annihilation and Compton scattering. This energy range has a great diversity of astrophysical objects and phenomena: nucleosynthesis, nova and supernova explosions, the interstellar medium, sources of cosmic-rays and cosmic-ray interactions, neutron stars, black holes, gamma-ray bursts, active galactic nuclei and the cosmic gamma-ray background. Not only do gamma-rays allow us to see deeper into these objects, but the bulk of the power radiated by them is often at gamma-ray energies.

The detection of photons in the energy band ranging from 10's of keV to a few MeV is dominated by the interaction between photons and detector material resulting in a measurement of the energy transfer during an interaction. The relevant interaction processes in this energy band are the photo-electric effect (below 100 – 300 keV), pair creation (above about 2 MeV), and Compton scattering at energies in between.

Detector materials are usually scintillators (NaI or CsI) where the energy loss of the photon locally ionizes the crystal, leading to scintillation flashes which are detected by photodiodes or photomultipliers. Solid-state semi-conductor detector materials are Ge (cooled to about 80 K) or CdTe (operated at room temperature). Here, the photon energy is correlated with the pulse height derived from an external bias voltage applied to the semi-conductor.

In the hard X-ray to soft gamma-ray regime, the typical signal-to-noise ratio is only a few percent. Therefore, typical exposure durations range from ~$10^3$ s up to ~$10^6$s. Usually, anti-coincidence systems are required to shield the detector from external background radiation such as photons from the diffuse cosmic background and from secondary photons and particles resulting from the interaction of primary cosmic-ray particles with the material of spacecraft and detectors. As sensitivity increases with the square root of the detection area, large detector arrays are desirable. However, this implies an increase in mass for both, detector array and the surrounding veto system. The latter might also become a source of secondary background radiation. Therefore, a careful optimization of the detector elements given the energy range and satellite constraints is mandatory.

High-energy photons can be focused through grazing incidence only up to energies of few 10's of keV. At higher energies, focusing is only possible in a limited way (for very narrow field of views and narrow energy bands) through a Laue diffraction lens. For wider field of view and broad energy ranges, other imaging techniques must be used for photon energies in the INTEGRAL range. These are the Compton imaging technique, where a photon is Compton scattered in one detector and photo-electrically absorbed in a second, spatially separated, detector, and the coded mask technique (see below for details).

# The INTEGRAL mission

## The scientific instruments

The **Inte**rnational **G**amma-**r**ay **A**strophysics **L**aboratory INTEGRAL (Fig. 1, Winkler et al., 2003) carries two main gamma-ray instruments, the spectrometer SPI optimized for the high-resolution gamma-ray line spectroscopy, and the imager IBIS optimized for high-angular resolution imaging. Two monitors, the soft X-ray monitor JEM-X and the optical monitor OMC, as well as a particle radiation monitor complement the scientific payload.

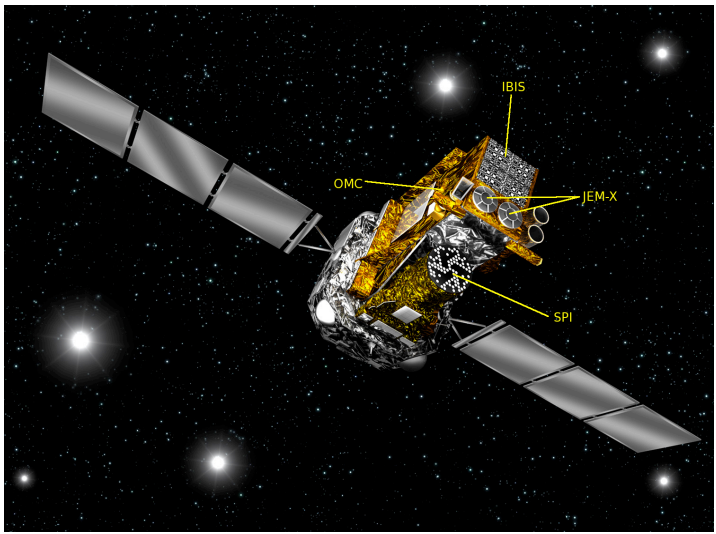

Figure 1: Artist impression of the INTEGRAL spacecraft. Dimensions are (5×2.8×3.2) m. The deployed solar panels are 16 m across. The mass is 4 t (at launch) including 2 t of payload, which consists of the imager IBIS, spectrometer SPI, and X-ray and optical monitors JEM-X and OMC. The coded aperture masks for SPI and IBIS are shown for illustration © ESA (C. Carreau).

The high-resolution gamma-ray spectrometer SPI (Vedrenne et al., 2003) consists of an array of 19 closely packed Ge detectors cooled by two pairs of Stirling Coolers to 80K. The detector array is surrounded by a Bismuth-Germanate (BGO) veto system. The imaging capabilities are obtained by placing a Tungsten coded aperture mask at a distance of 1.7 m above the Ge detection plane. SPI is providing the high-resolution spectroscopy of narrow lines and studies the large scale (> 1°) diffuse emission. The IBIS telescope (Ubertini et al., 2003) is the high angular resolution gamma-ray imager optimized for accurate point source imaging and for continuum and broad line spectroscopy. Its imaging system is based on two independent detector arrays optimized for low energies (ISGRI: 15-1000 keV, CdTe detectors) and high energies (PICsIT: 0.175-10.0 MeV, CsI detectors) surrounded by an active veto BGO system. A large Tungsten coded aperture mask (1 m²) is the imaging device placed at about 3.2 m above the detection plane (Table 1).

Table 1: Key parameters for SPI and IBIS

| **Parameter** | **SPI** | **IBIS** |
|---|---|---|
| Energy range | 20 keV - 8 MeV | 15 keV – 10 MeV |
| Detector | 19 Ge detectors [2] each (6 × 6 × 7) cm³, @ 80K | ISGRI: 16384 CdTe detectors, each (4 × 4 × 2) mm³, PICsIT: 4096 CsI detectors, each (8.55 × 8.55 × 30) mm³ |
| Detector area (cm²) | 500 | 2600 (ISGRI), 3000 (PICsIT) |
| Spectral resolution (FWHM) | 3 keV @ 1.33 MeV | 8 keV @ 100 keV |
| Field of view (fully coded) | 16° (corner to corner) | 8.3° × 8.0° |
| Angular resolution (FWHM) | 2.5° | 12' |
| Source location (radius) | < 1.3° (depending on source strength) | 1' for S/N = 30<br>3' for S/N = 10 |
| Continuum sensitivity | $5.0 \times 10^{-6}$ @ 100 keV<br>$1.3 \times 10^{-6}$ @ 1 MeV<br>(3σ detection in $10^6$ s, ΔE=E/2, photons cm$^{-2}$ s$^{-1}$ keV$^{-1}$) | $2.9 \times 10^{-6}$ @ 100 keV<br>$1.6 \times 10^{-6}$ @ 1 MeV<br>(3σ detection in $10^5$ s, ΔE=E/2, photons cm$^{-2}$ s$^{-1}$ keV$^{-1}$) |
| Line sensitivity (3σ, $10^6$ s, photons cm$^{-2}$ s$^{-1}$) | $4.7 \times 10^{-5}$ @ 100 keV<br>$3.4 \times 10^{-5}$ @ 1 MeV | $1.9 \times 10^{-5}$ @ 100 keV<br>$3.8 \times 10^{-4}$ @ 1 MeV |
| Absolute timing accuracy (3σ) | 130 μs | 61 μs |
| Mass (kg) | 1309 | 746 |
| Power [max/average] (W) | 385/110 | 240/208 |

The X-ray monitor JEM-X (Lund et al., 2003) provides X-ray spectra and imaging with arc-minute angular resolution in the 3 to 35 keV band in order to complement SPI and IBIS at lower ("softer") X-ray energies. JEM-X is a coded aperture instrument consisting of two identical, co-aligned telescopes. Each of the detectors has a sensitive area of 500 cm², and views the sky through its own coded aperture mask. The Optical Monitoring Camera OMC (Mas-Hesse et al., 2003) observes the optical emission (V, centered at 550 nm) from the prime targets of the high-energy instruments. OMC provides invaluable diagnostic information on the nature and the physics of the sources over a broad wavelength range. The OMC is based on a refractive optics with an aperture of 50 mm focused onto a large format CCD working in frame transfer mode (Table 2).

---

[2] *As of today, 15 out of 19 Ge detectors are operational: Ge detector #2 failed in December 2003, #17 in July 2004, #5 in February 2009, and #1 in May 2010.*

Table 2: Key parameters for JEM-X and OMC

| Parameter | JEM-X | OMC |
|---|---|---|
| Energy (wavelength) range | 3 keV – 35 keV | 500 nm - 600 nm |
| Detector | Microstrip Xe/$CH_4$-gas (@ 1.5 bar) | 50 mm f/3.1 refractor with CCD + V-filter |
| Detector area ($cm^2$) | 500 for each of the two JEM-X detectors | CCD: (2055 × 1056) pixels Imaging area: (1024 × 1024) |
| Spectral resolution (FWHM) | 3.6 keV @ 22 keV | -- |
| Field of view (fully coded) | 4.8° | 5.0° × 5.0° |
| Angular resolution (FWHM) | 3' | 23'' |
| 10σ source location (radius) | 1' (90% conf., 15σ source) | 2'' |
| Continuum sensitivity (3σ, $10^5$ s, ΔE=E/2, photons $cm^{-2}$ $s^{-1}$ $keV^{-1}$) | $8.5 \times 10^{-5}$ @ 6 keV $7.1 \times 10^{-5}$ @ 30 keV (for two JEM-X detectors combined) | -- |
| Line sensitivity (3σ, $10^5$ s, photons $cm^{-2}$ $s^{-1}$) | $1.7 \times 10^{-4}$ @ 6 keV $1.3 \times 10^{-4}$ @ 20 keV (for two JEM-X detectors combined) | -- |
| Limiting magnitude (10 × 200 s, 3σ) (50 × 200 s, 3σ) (100 × 200 s, 3σ) | -- | 18.1 $m_V$ 18.9 $m_V$ 19.3 $m_V$ |
| Sensitivity to variations | -- | Δ $m_V$ < 0.1 for $m_V$ < 16 |
| Absolute Timing accuracy | ~1 ms | > 3 s |
| Mass (kg) | 65 | 17 |
| Power [max/average] (W) | 50/37 | 26/17 |

SPI, IBIS and JEM-X share a common principle of operation: they are all coded aperture mask telescopes (Fig. 1). The coded mask technique (e.g. Caroli et al., 1987) is the key for imaging, which is all-important in separating and locating (faint) sources (Fig. 2). A mask consisting of an array of opaque and transparent (open) elements is placed between the source and the position sensitive detection plane. Every source within the field of view projects a shadowgram of the mask pattern on to the detector array. De-convolution techniques using the pre-defined mask pattern help to re-construct the location of the source. As such, a coded mask telescope works like a "pinhole" camera where the mask to detector distance defines the focal length, and the mask element size describes the angular resolution. Coded mask systems also provide near-perfect background subtraction, because for any particular source direction the detector pixels are split into two intermingled subsets, those capable of viewing the source through transparent (open) mask elements, and those for which the flux is blocked by opaque mask elements.

Effectively the latter subset provides a contemporaneous background measurement for the former, made under identical conditions.

All instruments are co-aligned with overlapping (fully coded) field of views ranging from 4.8º diameter (JEM-X), 5ºx5º (OMC), to 9ºx9º (IBIS) and 16º corner-to-corner (SPI), and they are operated simultaneously, providing the observer with data from all 4 instruments at any point in time.

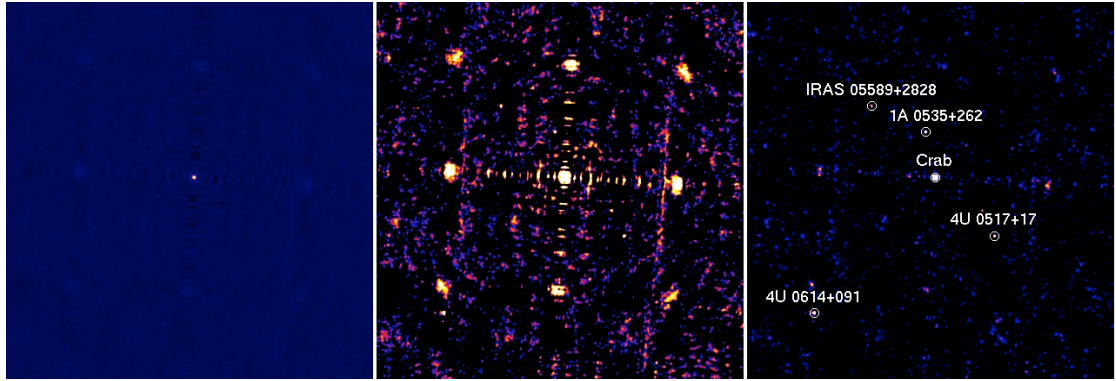

Figure 2: INTEGRAL IBIS image reconstruction in fields near bright sources as exemplified by the 18-40 keV image of the Crab region. Left: Image produced using the previous version of the image reconstruction algorithm. Middle: Around bright sources such as the Crab, the earlier algorithm results in "ghosts", i.e., artificial sources on a symmetric pattern around the real source. This image is identical to the one shown on the left, but uses a false color mapping that emphasizes faint background structures. Right: Same region and color map as middle image, showing how the improved handling of "ghosts" in the most recent release of the INTEGRAL software dramatically improves the image reconstruction and sensitivity (from the IBIS internal report for the ESA INTEGRAL Mission Extension Operations Review, July 2010).

With its broad energy range, INTEGRAL implements an astronomical bridge between the (soft) X-ray missions XMM-Newton, Chandra, Suzaku, Swift, and NuSTAR (which is planned from the end of 2012) and the space-based high-energy gamma-ray facilities in the GeV regime such as Fermi, AGILE, and ground based gamma-ray telescopes for higher (~TeV) energies such as H.E.S.S, MAGIC and VERITAS.

INTEGRAL's key astronomical capabilities rest on fine spectroscopy with imaging and accurate positioning of celestial sources of gamma-ray emission. Fine spectroscopy over the entire energy range permits spectral features to be uniquely identified and line profiles to be measured for physical studies of the source region. The fine imaging capability of INTEGRAL within a large field of view permits the accurate location and hence identification of the gamma-ray emitting objects with counterparts at other wavelengths, enable extended regions to be distinguished from point sources and provide considerable serendipitous science which is very important for an observatory-class mission.

INTEGRAL will remain unique world-wide as an observatory providing these capabilities.

**The INTEGRAL mission implementation**

A four-stage PROTON rocket launched INTEGRAL from the Baikonur cosmodrome in Kazakhstan on 17 October 2002, with a total spacecraft launch mass of about 4 tons, including a payload mass of about 2 tons. The spacecraft (Jensen et al., 2003) has been designed by utilizing the service module of the XMM-Newton mission. The PROTON operated flawlessly and injected INTEGRAL precisely into its orbit (Eismont et al., 2003). After a nominal mission duration of 2.2 years, INTEGRAL is now being operated[3] in its so-called "extended scientific mission operations phase", which has been re-approved by ESA in Fall 2010 to last until 31 December 2014, subject to a usual technical status review at the end of 2012. The on-board consumables (fuel, power) technically allow an extension of the operational lifetime beyond 2014. Further requests to extend the mission duration are being planned.

The orbit is characterized by a high perigee in order to provide long periods of uninterrupted observations with nearly constant background and away from the trapped radiation (electron and proton radiation belts) near the Earth. The orbital parameters at the beginning of the mission were: orbital period of 72 hours with an inclination of 52.2 degrees, a height of perigee of 9,050 km and a height of apogee of 153,657 km. Gravitational disturbances by Sun and Moon let inclination and heights of perigee and apogee vary in a quasi-periodic pattern: the perigee height for example will drop to about 3000 km early in 2012 followed by a raise to 10.000 km three years later.

Owing to background radiation effects in the high-energy detectors, scientific observations are carried out while the satellite is above a nominal altitude of typically 40,000 to 60,000 km. This means that most (~ 81 %) of the time spent in the 3-day orbit provided by the PROTON can be used for scientific observations, that is about 210 ks per revolution. An on-board particle radiation monitor allows the continuous assessment of the radiation environment local to the spacecraft.

Scientific and engineering telemetry data are downlinked 24 hours/day in real-time, as there is no on-board telemetry storage. The spacecraft is being operated using the ESA/ESOC facilities in Darmstadt/Germany. Scientific planning and operations (Much et al., 2003) including the scheduling of Target of Opportunity observations are being performed at ESA/ESAC near Madrid/Spain, in close consultation with the Project Scientist located at ESA/ESTEC (Noordwijk/The Netherlands). The processing of scientific data, quick-look analysis and archiving is provided at the INTEGRAL Science Data Centre (Courvoisier et al., 2003) in Versoix/Switzerland.

---

[3] An extensive description of the INTEGRAL mission, scientific payload, calibrations, science operations and initial "first-light" science results can be found in a special issue of A&A, Vol 411, No 1, November III (2003).

From the early beginning in 1989, the design and development of INTEGRAL was that of an observatory to be used by the scientific community at large. The observing programme, established through annual Announcements of Opportunity, is characterized by deep (>1 Ms) and often multi-year Key Programme observations, and, in contrast, also by flexible Target of Opportunity observations, enabling INTEGRAL to react to transient events of astrophysical significance in the strongly variable gamma-ray sky. The observing time for AO-8 observations (being executed in 2011) remains over-subscribed (the value of 3.8 is comparable to the over-subscription of ~Ms-long Key Programmes of other high-energy astrophysics missions). To make the best possible scientific use of the entire field of view (~100 □°), the community can also propose for data rights on individual point sources or extended sky regions, which are located in the field of view of observing programmes, but not included in their original science goals. This approach allows the most efficient and timely exploitation of the entire field of view data by various scientists. On average, ten scientific teams work on exploiting the science of each observation. INTEGRAL observations have resulted, so far, in more than 1450 publications[4], including 620 refereed papers, until June 2011.

Also, at least 74 PhD theses related to INTEGRAL science have been completed since launch until the end of 2010, and many more are on-going.

## Scientific Achievements and Future Prospects

Key science areas of INTEGRAL are:

- Studies of nucleosynthesis through gamma-ray lines from elements formed in supernovae,
- Studies of positron production and annihilation,
- Studies of the physics of emission mechanisms of white dwarfs, neutron stars, and black holes and associated transient phenomena
- Deep surveys for supermassive black holes in Active Galactic Nuclei, and
- Gamma-ray burst studies.

---

[4] *http://www.rssd.esa.int/index.php?project=INTEGRAL&page=Publications*

In the following, we give a summary of INTEGRAL's past achievements and the expected future science in the above areas. This will demonstrate, that INTEGRAL is extremely well suited for studying the physics of the high-energy universe, and future INTEGRAL observations are excellent opportunities to obtain answers for a wide variety of current questions in astrophysics.

**Nucleosynthesis, supernovae, and galactic structure**

INTEGRAL's capability for high-resolution gamma-ray line spectroscopy allows unique measurements of characteristic gamma-ray lines from the annihilation of positrons (511 keV), from the decays of radioactive isotopes $^{26}$Al, $^{60}$Fe, $^{44}$Ti (which all have been measured by INTEGRAL, e.g. Fig. 7), and others such as the $^{56}$Ni and $^{57}$Ni decay chains, $^{22}$Na and $^{7}$Be, which are still awaiting the opportunity of a sufficiently-nearby source event. Radioactive decays signify the existence of atomic nuclei, which must have been freshly produced in cosmic sites of nucleosynthesis. Such sites are supernovae of different types, interiors of rapidly evolving very massive stars, and novae. For the characteristic emission due to annihilation of positrons in interstellar space, the relation to their sources is less direct, and a subject of intense study.

Annihilation of positrons in interstellar space provides the brightest gamma-ray line signal, and thus sufficient data for detailed spectroscopy and its astrophysical exploitation. The precision spectroscopy of annihilation gamma-ray emission had been able to exploit line-to-continuum ratio and details of the line shape, in order to constrain tightly the nature of the interstellar medium where the annihilation actually occurs. It remains unclear, how tight the spatial connection between annihilation sites and positron production is, i.e. how (relativistic) positrons propagate between production and annihilation. Positron sources are a variety, from supernova and nova radio-activities through pulsars and microquasars and potentially even decay of dark matter.

The second-brightest gamma-ray line is the one from $^{26}$Al decay (1809 keV, decay lifetime 1.04 My), and INTEGRAL's deep exposures, thanks to its long mission duration, also raises the $^{26}$Al signal to a new quality level for astrophysical exploitation: Different source regions along the plane of the Galaxy can be separated (Fig. 4), such that their $^{26}$Al gamma-ray line intensity and shape can be evaluated for source yields, stellar-group ages, and for kinematic imprints from bulk motion or turbulence through the Doppler effect. In the inner Galaxy, shifts of the line centroid have been traced to vary systematically with Galactic longitude (Diehl et al., 2006, and Fig. 5); these are attributed to the differential large-scale rotation of source objects and their surrounding gas about the center of the Galaxy. With this measurement, INTEGRAL/SPI provides for the first time a kinematic trace of the hot and tenuous interstellar gas near its massive stellar sources. Compared to the longitude-velocity measurements for other (lower-mass) stars and for molecular clouds, the nucleosynthesis-enriched gas appears to move at significantly higher velocities. The penetrating gamma-rays help to trace the morphology of matter towards the inner Galaxy with spiral arms merging into a central bar near longitudes ±20°. Measurement of $^{60}$Fe radioactive-decay gamma-rays from the same massive star sources which provide the $^{26}$Al signal then allowed to derive a valuable constraint for the complex interiors of massive stars in their late stages: As these two radioisotopes result from different nuclear burning reactions in different regions of the star, models of stellar structure and evolution which cannot consistently reproduce the measured ratio can safely be discarded.

For individual sources of radioactive isotopes, the short-lived decays of $^{44}$Ti and $^{56}$Ni/$^{57}$Ni isotopes hold the prospect of unique messengers from the otherwise obscured supernova interiors, thus potentially helping a clarification and understanding of the still poorly known physical processes launching supernova and/or nova explosions of different types. Only the Cas A supernova remnant through its $^{44}$Ti ($\tau$~89 y) decay was bright enough to be seen by several hard X-ray and gamma-ray telescopes including INTEGRAL (Renaud et al 2006a). Gamma-ray spectroscopy of a SNIa would allow a unique "calibration" of its interior energy source, radioactive decay of $^{56}$Ni, helping to understand the physical origin of supernova brightness, which is a foundation of our claim for the existence of dark energy. A sufficiently nearby SNIa seems overdue, with an estimated event rate of about one to two per 5 years closer than 5 Mpc, but luck is part of science here.

In the following, we provide more detail on each of these science issues separately.

**Electron-positron annihilation emission**

The first large-scale sky map at 511 keV gamma-rays was produced by INTEGRAL (Knödlseder et al., 2005). The origin of the positrons producing this line is a 40-year old mystery. The physical channels capable of producing positrons range from conventional nucleosynthesis up to the decay of dark matter particles (see Prantzos et al., 2010 for a review). Sources, which are plausible producers of interstellar positrons include nucleosynthesis sources (through radioisotopes on the proton-rich side of the isotopes' valley of stability, i.e. mostly hydrogen-burning sites and supernovae), and sources ejecting relativistic plasma, such as accreting binaries or pulsars. More exotic or spectacular candidate sources would be gamma-ray bursters and the supermassive black hole in the center of our Galaxy, or the annihilation of dark matter particles.

The morphology of the galactic distribution of the 511 keV emission as measured by INTEGRAL (Fig. 3) severely constrains the number of possible models, and moreover presents a puzzle to astrophysics with respect to transport and slowing-down of positrons in interstellar space near sources and possibly in the Galaxy's halo. INTEGRAL can measure the detailed inner-Galaxy peak location and spatial structures in the distribution of the emission at 511 keV (Fig. 3) (Weidenspointner et al., 2008; Skinner et al., 2008), and determine the brightness of positron annihilation in the Galactic disk. Precision measurements of the line shape and positronium fraction identify the gas properties of the annihilation site (Churazov et al., 2005, 2011). This is key information for the determination of the origin of the positrons (Jean et al., 2009), and achievable within the extended INTEGRAL mission.

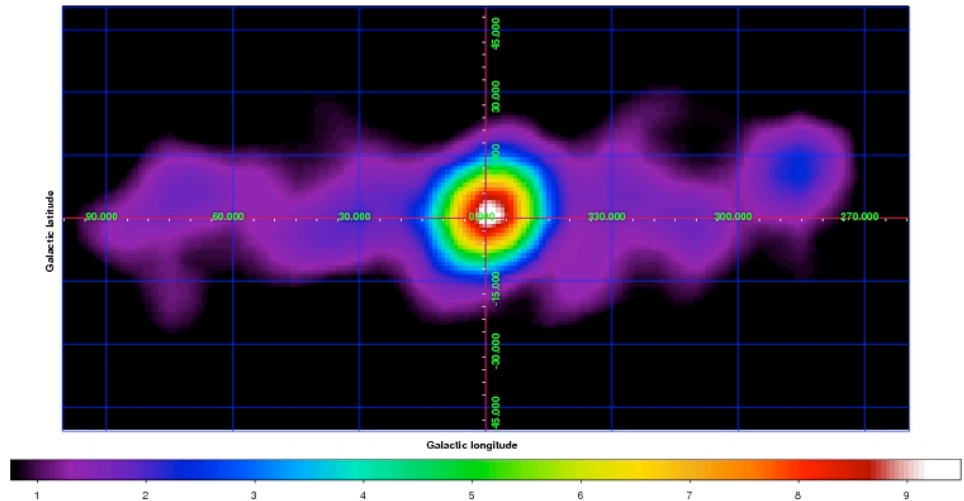

Figure 3: Spatial distribution of the 511 keV positron annihilation emission in the Galactic Centre. Long INTEGRAL exposures show a dominant central bulge and surprisingly weak emission from the disk. Although an earlier analysis was interpreted in terms of an asymmetry in the disk (Weidenspointer et al. 2008), a more recent one is shown here, as a significance map of the electron-positron annihilation emission in the 508 to 514 keV range, using data from six years of observation. It uses an alternative background modeling technique (Bouchet et al. 2010), suggesting that if any asymmetry is present, it is more likely associated with an offset of the bulge. Clearly, more observations are needed to understand the precise morphology and so provide key clues as to the origin of the positrons and how and where they annihilate (Figure adapted from Bouchet et al. 2010).

Both, cosmic-ray detectors near Earth, and the Fermi satellite, have measured positrons directly at GeV energies and above. Therefore, the existence of positrons in the plane of the Galaxy from these observations as well as from the observed decay of $^{26}$Al is proven, but, why is the annihilation so bright in the inner Galaxy? The answer may lie in the predominance of a source which is considered less plausible or even unknown at present; while dark matter annihilation or decay appears more exotic, the large-scale spatial distribution of accreting binaries with more-efficient plasma ejection may be revealed through annihilation gamma-rays, or the environment of the Galaxy-s supermassive black hole may provide cosmic-ray transport conditions very different from what we infer from large-scale galactic models. But long-distance propagation of positrons away from their sources may be possible through conspirations of magnetic-field structures in source regions with chimney connections towards the galactic halo and large-scale fields in the halo itself. In any case, deciphering the morphology of annihilation emission will result in a significant astrophysical lesson, including aspects of galactic structure, relativistic plasma sources, and astroparticle physics.

Through the capability to measure (red-shifted) 511 keV line emission, INTEGRAL provides unique diagnostics to probe black holes in X-ray transient outbursts.

**Diagnostics of past supernovae and supernova remnants**

Gamma-ray lines arise from radioactive decays of specific trace isotopes, thus give information on elements freshly created by nucleosynthesis processes. INTEGRAL has observed gamma-ray line emission from the long-lived (~1 My) isotopes $^{26}$Al and $^{60}$Fe, as well as from short-lived $^{44}$Ti.

Gamma-rays from $^{26}$Al sources along the plane of the Galaxy (Fig. 4) appear Doppler-shifted due to large-scale galactic rotation, directly proving its Galaxy-wide origin (Fig. 5), and allowing - using the measured total mass of $^{26}$Al - an independent determination of the galactic supernova rate due to core-collapse events terminating the evolution for all massive stars (Diehl et al., 2006). INTEGRAL can now disentangle $^{26}$Al source regions along the line of sight, and specifically measure $^{26}$Al yields from massive stars in associations (e.g. the Cygnus region (Martin et al., 2010), the Carina region, the Scorpius-Centaurus nearby groups (Diehl et al., 2010), and the Orion region (Voss et al., 2010)). Line shape information will be sufficient to constrain bulk motion down to the 100 km s$^{-1}$ range.

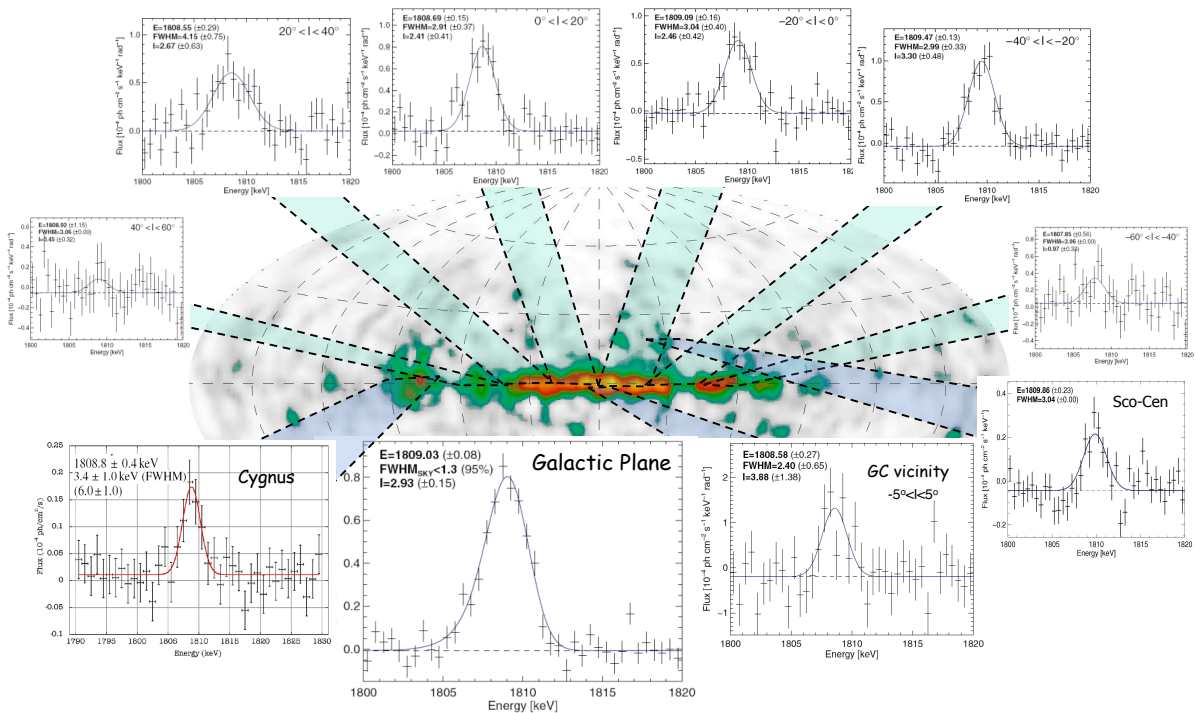

Figure 4: $^{26}$Al spectra along the galactic plane. The $^{26}$Al line traces nucleosynthesis in the Galaxy during the past ~$10^6$ years. Its emission can now be spatially resolved. This allows a comparison of the observed flux against nucleosynthesis models for massive stars and their supernovae in specific star formation regions (e.g. Martin et al. 2010).

INTEGRAL has applied its high spectral resolution for tracing the systematic Doppler shifts of the $^{26}$Al line centroid due to Doppler shifts from large-scale Galactic rotation (Kretschmer et al., 2011; see the longitude-velocity diagram for $^{26}$Al, Fig. 5). These results now constrain the dynamics of the hot interstellar medium in the inner Galaxy, through penetrating gamma-rays. It appears that the observed massive-star ejecta display higher velocities than what we know from their parental molecular gas. This measurement of gas kinematics along the Galaxy's central kpc regions may reveal how spiral arms connect to the Galaxy's inner bar, and eventually feed accretion towards the supermassive black hole.

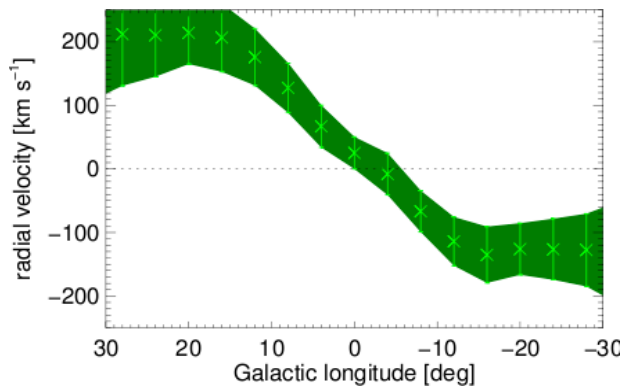

Figure 5: Doppler shift of the $^{26}$Al gamma-ray line at 1.8 MeV due to large-scale galactic rotation. First indicated in $^{26}$Al INTEGRAL data from the first 2 years of observations (Diehl et al. 2006), now, as shown in the figure, the line shift is strongly significant using data from 5.5 years of observations (Kretschmer et al. 2011).

Measurement of $^{60}$Fe radioactive-decay gamma-rays at sufficient significance (~5σ) and from the same massive star sources which provide the $^{26}$Al signal then allowed to derive a valuable constraint for the complex interiors of massive stars in their late stages (Wang et al., 2007): As these two radioisotopes result from different nuclear burning reactions in different regions of the star, models of stellar structure and evolution which cannot consistently reproduce the measured ratio can safely be discarded. INTEGRAL will finally be able to test whether the spatial distribution of the $^{60}$Fe line emission is consistent with that of the $^{26}$Al line emission.

$^{44}$Ti lines at 67.9 keV and 78.4 keV were detected from the young supernova remnant Cas A, clearly showing that this remnant originated from a (very) massive star. The non-detection of the 1157 keV line from the same decay chain places a lower velocity limit of 500 km s$^{-1}$ on ejected $^{44}$Ti (Martin et al., 2010).

In general, the measurement of specific amounts of radioisotopes from nucleosynthesis sources provide an overall validation or even calibration of nuclear reaction rates involved in their synthesis; as these rates often involve nuclear reactions which cannot be measured in terrestrial nuclear physics experiments directly, this constrains theories of the atomic nucleus and nuclear forces; contrary to common belief, these are far from being understood, in particular for isotopes heavier than e.g. Carbon, i.e. most of the cosmic 'metals'.

This is most convincing if individual/specific nucleosynthesis sources are involved. Thus, in its long-duration mission INTEGRAL will set a crucial constraint on the inner regions of core-collapse supernovae through the $^{44}$Ti gamma-ray line emission expected from these supernovae (The et al., 2006). Cas A, and possibly G1.9+0.3 (Borkowski et al., 2010) have been identified as such sources, previously unseen young SNR still embedded and occulted by molecular clouds (Renaud et al., 2006) are likely to be revealed through the homogeneous exposure of the galactic plane according to models. These results will guide future NuSTAR fine-imaging observations of $^{44}$Ti for specific SNR, and constrain core-collapse supernova models.

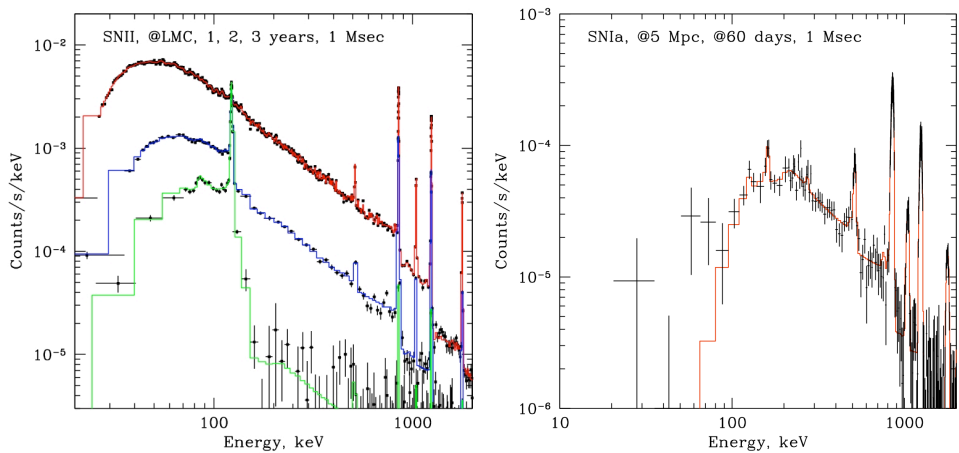

Figure 6: Predicted INTEGRAL/SPI spectra of nearby Supernovae. Left panel: SN Ia at a distance of 5 Mpc, 60 days after onset, 1 Ms integration. Model assumptions include: $M_{total}$ = 1.4 $M_{solar}$, with 0.7 $M_{solar}$ ($^{56}$Ni), 0.02 $M_{solar}$ ($^{57}$Ni); $v_{expansion}$ ~ $10^4$ km/s, with uniform mixing. Prominent emission lines are at 847 keV and 1238 keV ($^{56}$Ni). Right panel: SN II in the LMC (distance = 50 kpc), at 1, 2, and 3 years after onset, respectively, 1 Ms integration. The model parameters are similar as for SN 1987A. Prominent emission lines are at 122 keV ($^{57}$Ni), 847 keV and 1238 keV ($^{56}$Ni) (Figures produced by E. Churazov, priv. comm. 2010).

$^{56}$Ni decay provides the energy for supernova light of all supernova types, and also characteristic gamma-rays. Only gamma-rays <u>directly</u> measure the radioactivity which powers SN light, all other observables depend on models of the exploding star and its radiation transfer. Thus, INTEGRAL remains to be the premier mission to study supernova explosion mechanisms

through their nucleosynthesis and penetrating gamma-rays. A nearby exploding star event, either a supernova type Ia at a few Mpc, or a core collapse supernova in our own Galaxy or in nearby galaxies, would be hallmark events to prove or alter our models for the still unclear physics underlying those supernova, such as SN 1987A in the LMC (Fig. 6) had been. There is a realistic chance of such an event during the extended mission, and the measurement of gamma-rays from $^{56}$Ni decay with high spectral resolution (Fig. 6) could teach us more about supernova physics from one such object than from all previous ones combined.

Similarly, INTEGRAL's legacy with respect to nova nucleosynthesis will be based on tighter limits on $^{22}$Na and $^{7}$Be line gamma-rays, which have been predicted to be seen with INTEGRAL for novae if sufficiently nearby (<kpc; Hernanz & José, 2006).

In the early part of the mission, SPI's contribution was primarily the *detection* of gamma-ray lines (see Fig. 7). In the future, cumulative exposure will allow above-mentioned detailed studies of spectral line shapes and strengths and further studies of the spatial distributions to provide quantitative inputs for astrophysical models for element production, taking advantage of dedicated ultra-deep observations, complemented by lower cosmic-ray induced instrumental background.

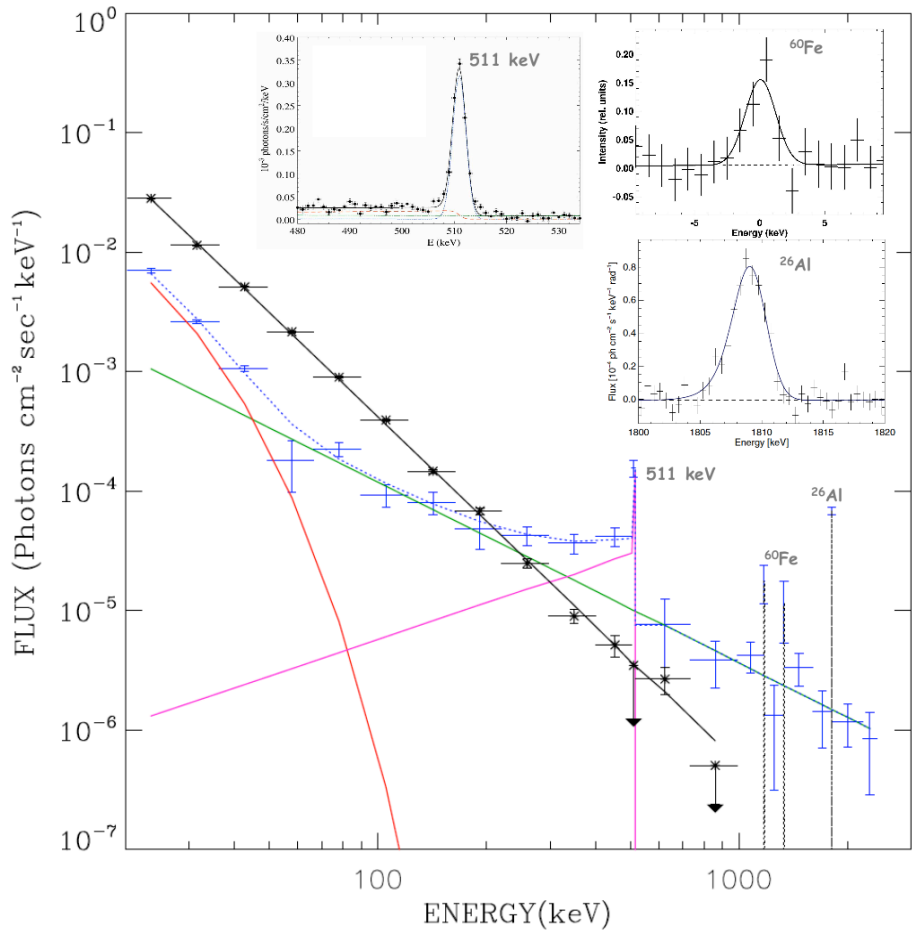

Figure 7: INTEGRAL spectra from the central galactic radian. Black: total emission of resolved point sources; Blue: total diffuse emission; Magenta: 511 keV annihilation radiation (positronium continuum and line emission); Green: emission from the galactic ridge resulting from interstellar particle interaction; Red: accreting magnetic white dwarfs (Bouchet et al. 2011, to be submitted). The vertical dotted lines indicate the narrow lines. Inserts show the fine spectroscopy analysis of selected lines. Note, that the insert of the $^{60}$Fe spectrum is a combined spectrum using the two $^{60}$Fe lines at 1173 and 1332 keV, respectively (from Jean et al. 2006; Wang et al. 2007; Wang et al. 2009).

## Compact galactic sources: introduction

Accreting compact objects such as stellar mass black holes and neutron stars exhibit broad-band spectra from the radio range up to several 100keV. In the X- and gamma-rays, these spectra are often produced by Comptonization, and therefore characterized by power law spectra with an exponential cutoff at a few 100keV. Above this energy, observations with SPI have allowed the detection of hard tails caused by emission from non-thermally distributed electrons, probably from electrons accelerated by shocks, e.g., in jets. Because X-ray binaries exhibit strong variability at all timescales and because the spectra are very hard, the long and uninterrupted observations provided by INTEGRAL are crucial for characterizing the physics of these objects. Since X-ray binaries are concentrated in the galactic plane, towards the Galactic Center, and in the bulge, INTEGRAL's large field of view allows studies of a large number and variety of these objects (Fig. 8). In this review, we discuss four scientific areas where the hard X-ray coverage of INTEGRAL was especially beneficial:

1. The discovery of the most strongly absorbed X-ray binaries and Supergiant Fast X-ray Transients.

2. Gamma-ray emission from accreting magnetized neutron stars,

3. Long-term studies of accreting black holes, and

4. The detection of gamma-ray polarization in the Crab pulsar and Cygnus X-1.

## Survey science: strongly absorbed X-ray binaries and supergiant fast X-ray transients

Due to its unique capability to monitor the Galactic Centre and bulge area for long periods at a sensitivity level unreachable by other wide-field instruments, INTEGRAL is especially capable of discovering faint and/or short-lived phenomena. At the time of writing, INTEGRAL discovered in total more than 60 new hard X-ray binaries and it doubled the number of known high-mass X-ray binaries (Bird et al., 2010).

This capability was illustrated with the extensive studies of the class of Supergiant Fast X-ray Transients (SFXT). These objects are high-mass X-ray binaries that display short X-ray outbursts typically lasting only hours and with a low duty cycle (Ubertini et al., 2005; Sguera et al., 2006; Sidoli et al., 2007). At the time of writing, the total number of these sources was already comparable to that of persistent supergiant X-ray binaries, with new serendipitous discoveries routinely being made. What causes the short outbursts is still debated, mainly due to small source number statistics. There is agreement, however, that it is connected to the mode of stellar wind accretion (Sidoli et al., 2007; Grebenev & Sunyaev, 2007; Bozzo et al., 2008, Ducci et al., 2009). For example, in a recent model that is based on a study of 30 Ms of INTEGRAL monitoring of SFXTs, Ducci et al (2010) argue that the distribution of flare luminosities is in principle consistent with accretion from a clumpy stellar wind (see also Romano et al., 2010). The distribution of mass accretion rates inferred for these sources, however, differs from that obtained from Hα modelling of these systems. Ducci et al. (2011) argue that a combination of photo-ionization of the accreted material and inhibition of accretion by centrifugal forces leads to the formation of transient accretion disks which would explain these differences.

With the discovery of the class of strongly absorbed X-ray binaries in INTEGRAL observations (Walter et al., 2003; Revnivtsev et al., 2003; Lutovinov et al., 2005; Fig. 8), a large number of new compact objects were discovered. These objects are embedded in environments with very high column depths of $N_H > \sim 10^{24}$ cm$^{-2}$, such that these sources could not be identified earlier with soft X-ray instruments (Kuulkers, 2005; Ibarra et al., 2007; Barragan et al., 2009). More than half of these sources have supergiant companions, where a significant fraction of the absorption is due to the supergiant's stellar wind. Observations of several of the strongly absorptions show X-ray pulsations, indicating that the compact object is a neutron star (Walter & Zurita Heras, 2007; Rahoui et al., 2008; Chaty et al., 2008).

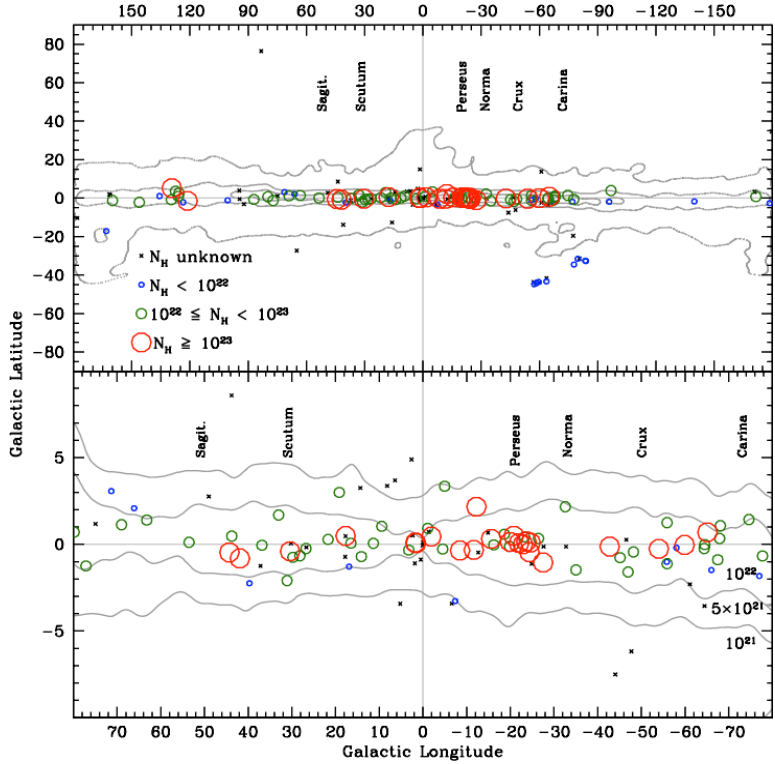

Figure 8: Spatial distribution (in Galactic co-ordinates) of all INTEGRAL/ISGRI-detected sources belonging to the Milky Way and Magellanic Clouds where the symbol size is proportional to the published column density ($n_H$) derived from an X-ray spectrum. The contours denote expected absorption levels through the Galaxy (Dickey & Lockman 1990). Courtesy A. Bodaghee (priv. comm. 2011).

**Neutron stars and black holes**

The long (~$10^6$ s) observing campaigns on neutron star and black hole X-ray binaries are especially suited to study the interplay between the different components of the accretion flow onto these objects such as the cold disk, hot Comptonizing plasma, and their synchrotron radiating jets or outflows (Rodriguez et al., 2008; Cadolle Bel et al., 2009; Prat et al., 2009).

INTEGRAL discovered hard tails extending to 200 - 300 keV in extremely magnetized neutron stars ("magnetars"), which show a complex dependence on the pulse phase. The observations (Fig. 9) of these anomalous X-ray pulsars (AXP) provided new constraints on the geometry and physics of the strongest magnetic fields in the Universe (Kuiper et al., 2004; Kuiper et al., 2006; den Hartog et al., 2008; den Hartog et al., 2008a; Mereghetti et al., 2009).

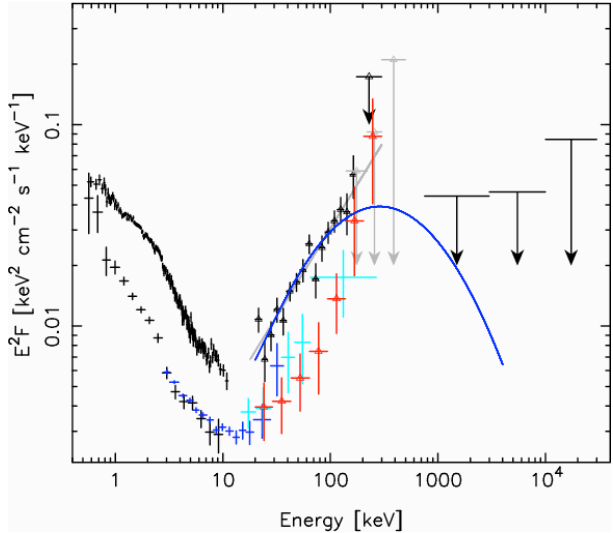

Figure 9: The luminous non-thermal emission from magnetars above 10 keV was discovered by INTEGRAL and still remains unexplained. Black data points show the total energy spectrum of 1RXS J1708-40 from XMM-Newton (<12 keV), INTEGRAL/IBIS (triangles), as well as upper limits from INTEGRAL/SPI (grey upper limits) and CGRO/COMPTEL. This spectrum is consistent with either a power law (grey) or a log-parabola (blue). The pulsed component of the emission is weaker and shown here using data from XMM-Newton (black), RXTE-PCA (blue), HEXTE (cyan), and INTEGRAL/IBIS (red). Credit: den Hartog et al., A&A **489**, 263, 2008, reproduced with permission © ESO.

Another 'flavour' of a strongly magnetized neutron star, or "magnetar" is the soft-gamma repeater (SGR). These SGR events differ from the majority of classical gamma-ray bursts (GRB) in that they are of much shorter duration and have relatively soft spectra. The recurrent sources can thus be considered as a distinct class of objects (see, e.g., Kouveliotou et al., 1993). INTEGRAL discovered persistent hard X-ray emission from the soft gamma-ray repeater SGR 1900+14 while it was in a quiescent state (the last bursts from this source were observed in 2002). By comparing the broad-band spectra (1-100 keV) of the magnetars (SGR and AXP) observed by INTEGRAL, evidence was found for a different spectral behaviour of these two classes of sources: SGR spectra are soft at energies above 10 keV and hard below 10 keV, <u>in contrast</u> to spectra from AXP which show an opposite spectral shape. Therefore, the determination of a spectral cutoff at 0.1 MeV or at 1 MeV should discriminate between various magnetar models (Götz et al., 2006).

In neutron star systems with weaker magnetic fields ($B \sim 10^{12}$ G), cyclotron lines are the only direct way to determine the magnetic field strength of neutron stars (Schönherr et al., 2007 and references therein). At these magnetic field strengths, the motion of electrons around the magnetic field lines is quantized into Landau-levels. The inelastic scattering of hard photons from the accretion mound with these quantized electrons leads to the formation of cyclotron lines. The energy of these lines ($E_{cyc}$) is proportional to the magnetic field strength in the line forming region, $E_{cyc}$ = 12 keV ($B/10^{12}$ G), and can therefore be used to measure directly the B-field. INTEGRAL monitoring of spectral changes during the outbursts of neutron star binaries with cyclotron lines (Fig. 10) led to the discovery that the measured line strength is correlated with source luminosity. This result is interpreted as a change in height of the accretion column over the outburst, which is due to the changing mass accretion rate: At the high accretion rates during these transient outbursts, the characteristic height of the accretion column is the location of the radiative shock where the accreted material is decelerated to speeds below the local speed of sound (Becker & Wolff, 2007). As the mass accretion rate increases, the radiative pressure caused by photons produced in the accretion column increases. As a result, the radiative shock moves away from the neutron star's surface into regions of lower magnetic field, leading to a decrease in the energy of the cyclotron line at higher luminosities (Tsygankov et al., 2006; Mowlavi et al., 2006; Caballero et al., 2008).

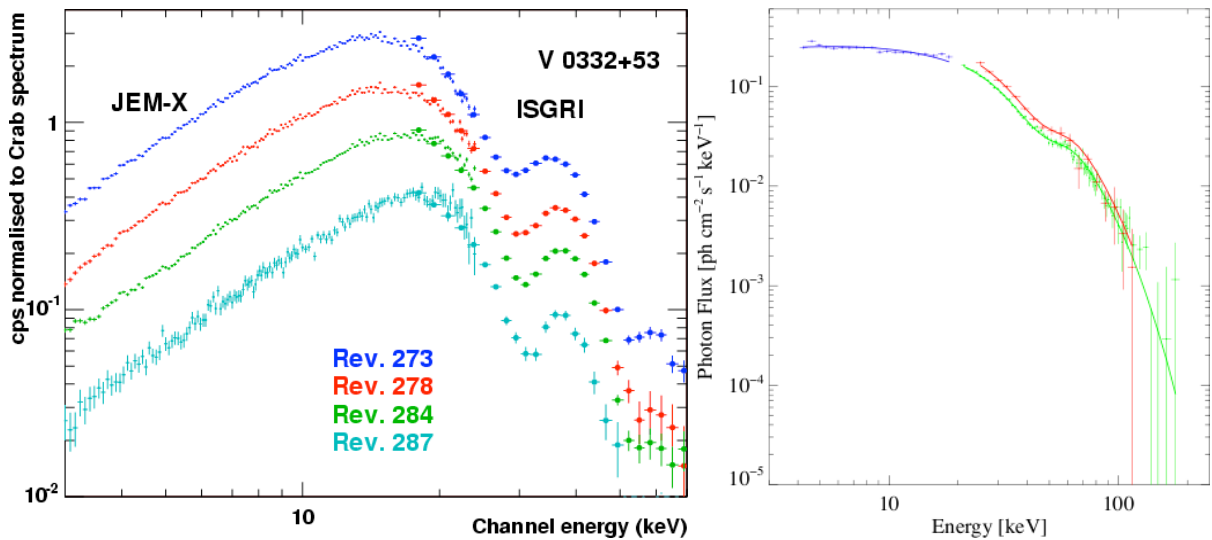

Figure 10: INTEGRAL detections of cyclotron lines in absorption.

Left panel: Spectra from the high-mass X-ray transient V0332+53 during the 2005 outburst decline. JEM-X and ISGRI spectra integrated, from upper to lower curves, during revolutions 273, 278, 284 and 287, respectively. The spectra have been normalized to the Crab spectrum of revolution 239. Credit: Mowlavi et al., A&A **451**, 187, 2006, reproduced with permission © ESO.

Right panel: INTEGRAL spectrum (JEM-X: blue, ISGRI: red, SPI: green) of the Be transient A0535+262 during its August 2009 outburst. The cyclotron line at ~ 45 keV is clearly visible in the ISGRI and SPI spectra (I. Caballero, priv. comm. 2011).

Other INTEGRAL detections on compact objects, which shall only be mentioned briefly here, included:

(i) hard emission from black holes up to 1 MeV (e.g. Caballero García et al., 2007)

(ii) non-thermal components >100 keV from neutron star low-mass X-ray binaries (Fiocchi et al., 2006; Paizis et al., 2009; Cocchi et al., 2010)

(iii) the Compton up-scattered radiation at >100 keV in accreting millisecond pulsars in outburst (Falanga et al., 2005; Falanga et al., 2007).

**X-ray polarization measurements**

The INTEGRAL/IBIS detector consists of two spatially separated arrays (ISGRI and PICsIT), therefore, it can be operated as a polarization sensitive Compton telescope (Forot et al., 2007). If a photon interacts with an electron in the upper detector array of IBIS (ISGRI) and is then absorbed in the lower IBIS detector array (PICsIT), it is possible, to reconstruct the energy of the incident photon as well as the scattering angle from both, the measurements of the location of interaction and of the deposited energy. The differential cross-section for Compton scattering depends on the polarization of the incident photon, hence, a measurement of the degree of polarization and polarization angle becomes possible. Similar measurements are also possible with SPI, where photons (as multi-site events) interact with more than one Ge detector.

INTEGRAL detected the degree and phase-dependence of polarization in the Crab nebula and pulsar in the hard X-ray to gamma-ray regime using IBIS (Forot et al., 2008) and SPI (Dean et al., 2008). Hitherto, polarization from Crab had been measured from the radio up to soft X-rays with photon energies up to 5.2 keV (Weisskopf et al., 1978). The degree of polarization and the polarization angle are comparable to that measured in the optical for the point source in the Crab nebula, indicating that the hard X-rays and gamma-rays from the Crab originate in the same region, close to the pulsar, as those seen in the optical wavelength range.

Recently, use of the IBIS Compton mode allowed detecting polarization in the hard tail of the black hole candidate Cygnus X-1 (Laurent et al., 2011). This is the second source outside the solar system only, in which polarization has been detected. While the X-rays below ~400 keV are not polarized in Cygnus X-1, which is consistent with an origin of the radiation in a Comptonizing plasma, X-rays in the hard tail (above 400 keV) of the source are strongly polarized. Consistent with some emission models for black holes (e.g., Markoff et al., 2005), this high degree of polarization is likely due to synchrotron radiation, which could be emitted, for example, in the base of the radio jet seen during the hard state of this system.

In the near future, very deep observations of further selected bright X-ray sources will allow further constraints of polarization, which will be complementary to similar polarization measurements in the soft X-rays, which are expected to become available with the launch of NASA's GEMS satellite after 2014.

## The centre of the Galaxy and galactic diffuse emission

INTEGRAL discovered persistent hard X-ray emission from the Centre of the Galaxy (Bélanger et al., 2004; Bélanger et al., 2006) (Fig. 11).

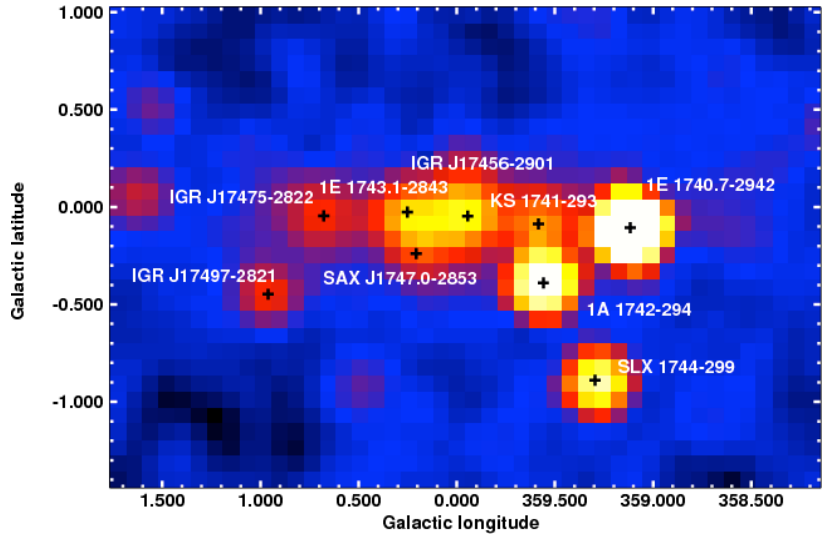

Figure 11: INTEGRAL view of the Galactic Centre in the 20-40 keV energy range, obtained from all IBIS observations of this region between 2003 and 2009, for a total exposure of 20 Ms. The main sources are indicated by black crosses. Most of them are X-ray binaries powered by accretion on neutron stars or stellar-mass black holes. The central source IGR J17456-2901 coincides with the position of Sgr A*. The position of IGR J17475-2822 is compatible with that of the Sgr B2 molecular cloud (from Terrier et al. 2010. Reproduced by permission of the AAS.)

This emission is not due to the hot plasma seen at lower energies with Chandra or XMM-Newton, and neither is it due to the central black hole, Sgr A* (Trap et al., 2010). INTEGRAL's discovery of hard X-rays from the giant molecular cloud Sgr B2, which are best interpreted as scattering of radiation emitted by Sgr A* more than 100 years ago (Revnivtsev et al., 2004), shows that this non-thermal emission likely traces the past activity of Sgr A*. The most compelling evidence of such reflection is the INTEGRAL discovery of Sgr B2 fading in hard X-rays (Terrier et al., 2010, see Fig. 12).

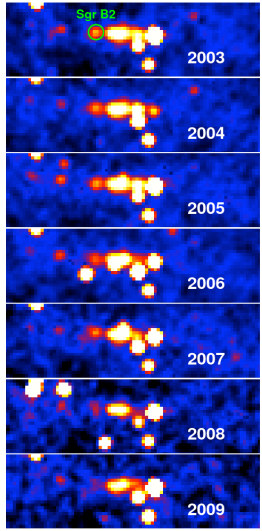

Figure 12: INTEGRAL IBIS 20 – 60 keV images of the Galactic Centre (7º × 2º) from 2003 until 2009. Most of the sources are X-ray binaries and display strong time variability. The green circle shown in the top panel corresponds to the position of the molecular cloud Sgr B2. The associated hard X-ray source, IGR J17475-2822 (see also Fig. 11), shows a clear decline in flux during the 7 years of monitoring (from Terrier et al. 2010. Reproduced by permission of the AAS.)

Along with XMM-Newton measurements of the Fe K-line variability from other clouds of this region (Ponti et al., 2010), this discovery allowed both the strength to be constrained and the duration of this past activity to be limited. These results established that the behaviour of Sgr A* resembles that of a low-luminosity active galactic nucleus and it might become brighter again in the future. During the continued monitoring of the current activity of Sgr A*, INTEGRAL surveys of the Galactic Centre will also allow the past activity of the closest super-massive black hole to be monitored by detecting the Compton echo of its outburst radiation as it propagates through the molecular clouds of the region.

Outside of the Galactic Centre, INTEGRAL discovered that accreting white dwarfs are responsible for a significant fraction of the originally apparently "diffuse" galactic ridge emission (Krivonos et al., 2007, see also Fig. 7). With ultra-deep (~10 Ms) observations of the galactic bulge and arms (Fig. 13), INTEGRAL will continue to resolve the galactic ridge emission, to separate the point source population from the diffuse emission, and finally detect the predicted low-luminosity source population. Coverage of the extended plane of the Galaxy with high exposure is essential for this science but has not yet been obtained. Disentangling the source contribution from extended diffuse emission provides important information for understanding the gas and energy content of the Galaxy as a whole and its interstellar medium.

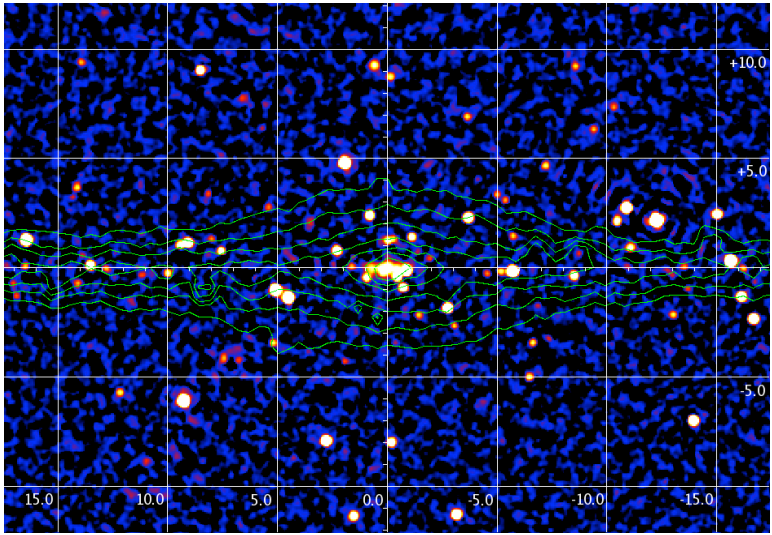

Figure 13: The best existing image of the central part of the Galaxy as obtained with INTEGRAL/IBIS (17-60 keV). The color-scale of the image was chosen such that the statistical noise is seen as black-blue regions and red-yellow-white circles are detected objects. Green curves are iso-contours of surface brightness in the infrared from COBE/DIRBE, tracing the surface density of stars in the Galaxy. The maximum sensitivity of the map is within 5º to 10º of the Galactic Centre. Credit: Krivonos et al., A&A **519**, A107, 2010, reproduced with permission © ESO.

Through these continued deep observations, INTEGRAL will also allow new and rare phenomena in the Galaxy to be uncovered, such as outbursts of supergiant fast X-ray transients, magnetars, and accreting millisecond pulsars, which cannot be discovered with other missions such as Swift or Fermi. Only INTEGRAL can effectively detect their outbursts and answer basic questions such as: What is the physics behind short outbursts? What are the outburst duty cycles and are they periodic? Is there persistent emission outside outbursts? Observations of further cyclotron lines with INTEGRAL will also allow the number of neutron star systems with directly measured magnetic fields to be extended (only ~20 are known to date).

Of special interest are also binary systems emitting at GeV and TeV energies. So far, Fermi, H.E.S.S., AGILE, and MAGIC have observed only four binaries emitting radiation up to GeV/TeV energies. An increase in source population is expected as H.E.S.S.-II and MAGIC-2 become operational, Fermi's survey gets deeper, and CTA is built. The nature of these systems and the origin of the extreme gamma-ray emission are under debate (Ubertini et al., 2009). For these systems, as well as for supernova remnants, pulsar wind nebulae and their pulsars, simultaneous hard X-ray/soft gamma-ray coverage in the gap between X-rays and GeV/TeV is necessary to understand the non-thermal emission and related particle acceleration processes in GeV/TeV systems.

## Supermassive black holes, the cosmic X-ray background, and cosmology

The hard X-ray band, from 15 keV to 200-300 keV is a critically important region of the astrophysical spectrum for the study of active galactic nuclei (AGN). In this band, an unusually broad range of astrophysical processes occurs. This is the energy domain, where the fundamental change from primarily thermal to non-thermal phenomena occurs and where the effects of inter-galactic absorption are drastically reduced. Furthermore, most of the extreme astrophysical behavior shown by cosmic sources is taking place in this energy range.

To illustrate the power of hard X-ray surveys for understanding AGN, we mention two examples where the hard X-ray/soft gamma-ray observations are crucial: to unveil the nature, the number and the cosmic evolution of absorbed sources, and to explore spectral features present in AGN: Figure 14 compares the broad band spectrum of 3C 273, a very bright AGN, known to be a prototype of non-absorbed extragalactic sources, to that of MKN 3, which is instead a heavily absorbed, so-called "Compton-thick" source in view of the large amount of material absorbing most of the radiation which is emitted by its central nucleus. While the two objects are similar at energies above 20 keV, i.e. the central nucleus has the same emitting power, the two are remarkably different at lower energies where the emission from MKN 3 is significantly depleted by the high absorbing column density of a few $10^{24}$ cm$^{-2}$ (Cappi et al., 1999). Extragalactic objects with such a high absorption can only be found and studied by means of wide field of view instruments capable to perform hard X-ray surveys/observations.

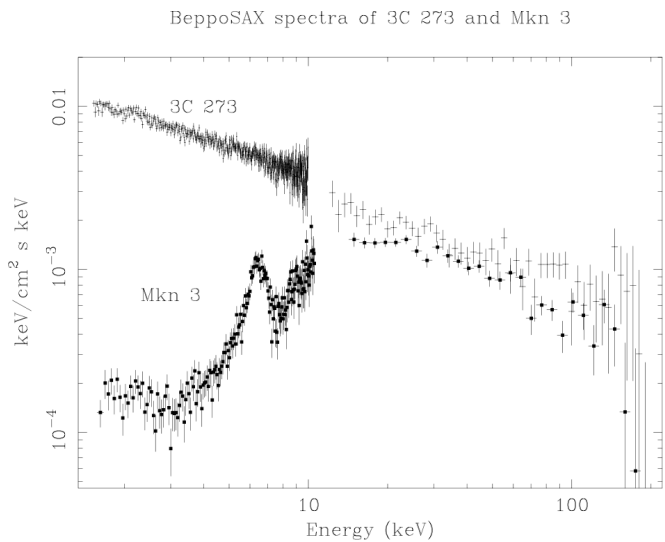

Figure 14: The broad-band spectra of Mkn 3 and 3C 273 (L. Bassani, priv. comm. 2011). As can bee seen, the behavior of the two AGNs is similar at high energies but very different at E < 20 keV. The high energies surveys are a powerful tool to discriminate between Compton thick and unabsorbed AGNs.

Another reason for studying AGN at high energies is to obtain broad band spectral information on the various classes of AGN: for Seyfert galaxies, this provides information on spectral features like the reflection hump, due to nuclear emission which is Compton scattered by the dusty torus surrounding the AGN, and the high energy cut-off, that can only be studied with highly sensitive instruments at energy above 10 - 20 keV (for a comprehensive review of AGN spectral features, see Mushotzky et al., 1993). Finally, deep observations of high- redshift Blazars in the range up to hundreds of keV opens a new window in their Spectral Energy Distibution (SED) allowing the study of extreme or peculiar objects with a nuclear mass of up to $10^{10}$ solar masses (Ghisellini 2009).

INTEGRAL plays a key role in our understanding of AGN covering the energy range from 10 keV up to 300 keV where the non-thermal processes are dominant. The IBIS all-sky catalogue (17-100 keV) comprises now more than 250 AGNs (Bird et al., 2010) of which only 9 were initially detected in the energy range 100-150 keV and a few between 150 and 300 keV (Bazzano et al., 2006). So far, the number of AGNs detected above 100 keV has more than doubled and is monotonically increasing with deep exposure outside of the galaxy plane. This is a confirmation that INTEGRAL is reaching a sensitivity good enough to firmly detect the high energy cut-off of at least the closest Seyferts and brightest Blazars (Panessa et al., 2011). Broad-band spectral analysis of more than 100 AGN has shown that their high-energy radiation is due to thermal Comptonisation of soft photons by a plasma surrounding the central black hole, which is mildly relativistic and has a low optical depth (Molina et al., 2009; Beckmann et al., 2009). The latest INTEGRAL catalogue (Bird et al., 2010) has significantly increased the number of these objects detected above 10 keV to more than 250, spanning a large range in redshift ($0 \leq z \leq 3.7$) and luminosities ($10^{42}$ -$10^{48}$ erg/s) (Fig. 15). Most of them are Seyferts type 1 and type 2 in almost equal percentage, while only a few of the brighter Blazars (BL Lac and flat spectrum radio QSOs) have been detected so far.

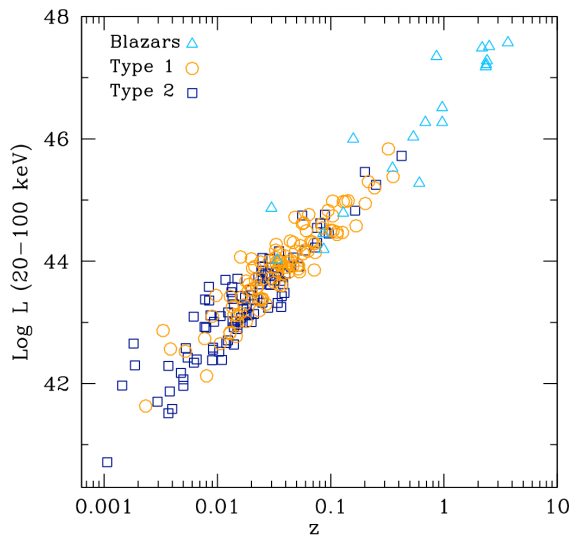

Figure 15: Hard X-ray luminosity versus redshift for the 258 AGN contained in the 4[th] INTEGRAL source survey (Bird et al. 2010). Figure adapted from Malizia et al. 2009b.

INTEGRAL is a key instrument in the study of heavily absorbed AGN: it has extensively probed the Compton thin regime ($N_H \leq 1.5 \times 10^{24}$ cm$^{-2}$) and it has discovered new Compton thick sources (Malizia et al., 2009a). INTEGRAL has revealed, that the percentage of absorbed sources ($N_H > 10^{22}$ cm$^2$) is ~60%, while the fraction of Compton thick objects ($N_H > 10^{24}$ cm$^2$) is closer to 10% (Sazonov et al., 2007; Paltani et al., 2008; Malizia et al., 2009b; Treister et al., 2009) in contrast to results obtained from optically selected samples. INTEGRAL has also shown a trend of a decreasing fraction of absorbed AGN with increasing hard X-ray luminosities. In the next 4 years, ultra-deep (≥10 Ms) observations will permit to disentangle whether this is a direct consequence of the evolution of the AGN luminosity function with redshift. We anticipate the number of high-energy detected AGN will double as every optical spectroscopic follow-up campaign finds new AGN among the unidentified IBIS sources. Most importantly, as the sky exposure deepens in the coming years, other types of high-energy AGN will emerge and their physics will be studied in INTEGRAL's hard X-ray/soft gamma-ray range for the first time, complemented below ~70 keV with NuSTAR's follow up observations.

INTEGRAL will also provide unique and essential observational coverage in the soft gamma-ray domain for different AGN classes in synergy with X-ray and gamma-ray observatories, ground based TeV telescopes, and radio-VLBI monitoring of their jets. Long term monitoring of the variability for several very bright AGN is needed to unveil the geometry and the physics of the accretion on the central engine in the non-thermal regime. Wide-band INTEGRAL spectra are and will be a unique tool to characterize the primary spectral shape of AGN, which is of paramount importance for synthesis models of the cosmic X-ray background (CXB), as they will provide key parameters such as power law index, cut-off energy and the strength of any reflection component. INTEGRAL has provided first direct comparison between the collective hard X-ray SED of local AGN and the CXB spectrum in the 3–300 keV energy range (Sazonov et al., 2008).

The main objective for the coming years will be the observation of a large sample of jet-dominated Blazars, jointly with Fermi, ground based gamma-ray facilities, and radio-VLBI. Such studies will provide an independent insight into the physics of jets with INTEGRAL's unique sensitivity achieved in deep extragalactic fields. Some of those Blazars have the most powerful jets, larger black hole masses and more luminous accretion disks: their spectral energy distribution has a Compton peak in the MeV to sub-MeV region (Bassani et al., 2007; Masetti et al., 2008).

The INTEGRAL discovery of such massive objects is shedding a new light on the co-evolution of black holes and galaxies and starts to probe the initial supermassive black hole (Della Ceca et al., 2009). Reaching an ultra deep exposure in the next few years is of paramount importance in order to discover hard X-ray emitting Blazars that will provide a new perspective into jet dominated sources thereby complementing and superseding the standard optical classification. INTEGRAL observations will allow a unique investigation of the recently suggested contribution of Blazars to the diffuse gamma-ray continuum in the 100-300 keV range, where that of Seyferts is decreasing exponentially.

Additional observations, using the Earth as a "blocking device" (see Churazov et al., 2008; Türler et al., 2010 for initial results), will be performed to determine the integrated hard X-ray background flux with INTEGRAL.

Observations within the hard X-ray range will add fundamental information to our overall knowledge of super-massive black hole activation, unification, and evolution and provide an invaluable database from which to select the main scientific targets for the forthcoming hard X-ray missions, such as NuSTAR and ASTRO-H. The INTEGRAL survey capability, utilizing its unique ~100 □° field of view will be complemented by NuSTAR's superior resolving power and sensitivity follow-up provided below 70 keV. In fact, NuSTAR (and ASTRO-H) would need, due to the small field of view of about 0.05 □°, about $10^6$ separate pointings to cover the entire sky.

The high-energy survey legacy of INTEGRAL and Swift/BAT will therefore remain unique for the next decades and the "decade long" unbiased INTEGRAL survey will be of seminal importance for extragalactic astronomers, comprising around 1500 sources by the end of 2014 (Fig. 16).

## Survey projection for 2014

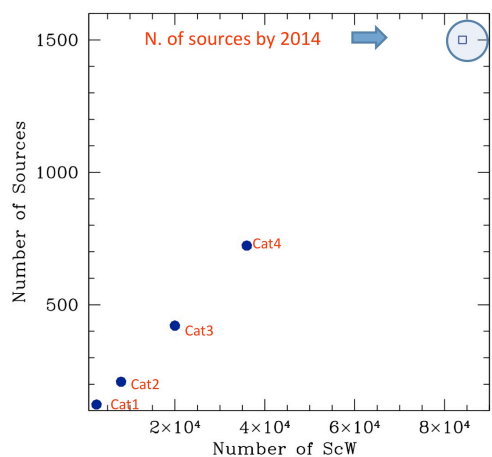

Figure 16: Expected number of sources catalogued by INTEGRAL until 2014 (1 Science Window [ScW] corresponds to an integration time of about 3600 s).

**Gamma-ray bursts**

INTEGRAL is one of the major facilities for studying Gamma-Ray Bursts (GRB) above 20 keV. Real time, arc-minute positions of GRB imaged with IBIS are distributed via the WWW at a rate of about 10 per year (75 to date), the weakest GRB with a fluence of ~$5\times10^{-8}$ erg cm$^{-2}$ (Vianello et al., 2009). Follow-up observations by Swift, XMM-Newton, and ground-based observatories have, at the time of writing, revealed 45 X-ray, optical, and/or radio counterparts to these events, with redshifts between z=0.105 and 3.793. The SPI ACS has played a key role in localizing and identifying two events, which are believed to be extragalactic giant magnetar flares from M81 and M31 (Frederiks et al., 2007; Mazets et al., 2008). As LIGO was operative at the time of the M31 event, neutron star-neutron star mergers could for the first time definitively be excluded for these events, whereas the data are consistent with the magnetar flare hypothesis (Abbott et al., 2008).

Compared to Swift, INTEGRAL with its unique sensitivity detects a larger number of faint GRB (Fig. 17). This is essential to investigate the fraction of GRB with long spectral lags, which appear to be a low-luminosity population distinct from the high-luminosity one and which is inferred to be local (Foley et al., 2008) - an important INTEGRAL discovery. This evidence will be confirmed in the next years almost doubling the sample, also in view of the increasing detector sensitivity due to lower cosmic-ray induced background while approaching the next solar maximum. Recently, a confirmation of the existence of sub-luminous GRB has been announced by MAXI (Serino et al., 2011), the Japan-USA experiment on the ISS, operational since 15 August 2009 (Kawai et al., 2011). In the first 16 months of scientific observations, MAXI has detected, in the energy range of 2-20 keV, an average of 12 GRB/y, versus the pre-launch prediction of 3.2 GRB/y. Eight out of 16 are weak GRBs, below the detection sensitivity of previous X-Ray satellites, and less sensitive than MAXI. Of the 16 GRBs detected only 5 are in coincidence with other detections. Of particular interest is GRB091230, a weak burst seen in coincidence with INTEGRAL. We expect MAXI to shed new light on the sub-luminous population of GRBs, so far detectable by INTEGRAL only.

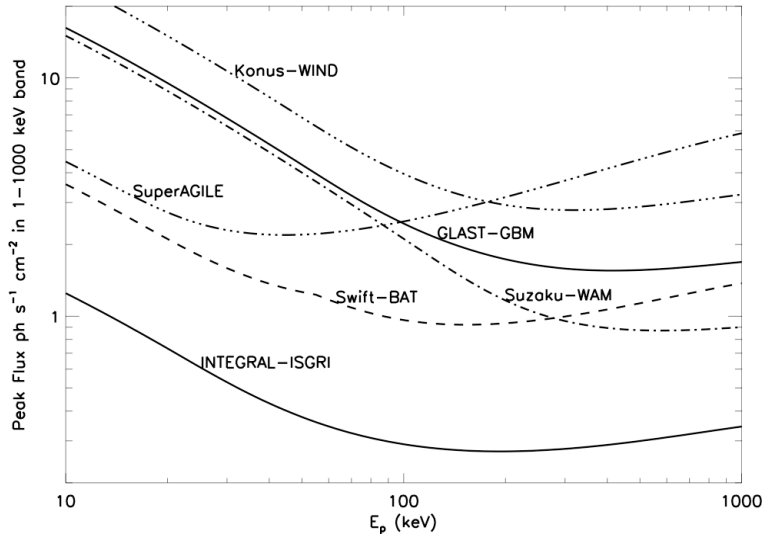

Figure 17: Sensitivity of current missions for the detection of faint GRB. Reprinted with permission from D. Band et al., AIP Conf. Proc. **1000**, 121-124 (2008), Gamma-Ray Bursts 2007: Proceedings of the Santa Fe Conference. © 2008, American Institute of Physics.

INTEGRAL will remain the only mission with the arc-minute imaging capability and wide band sensitivity necessary to measure prompt GRB emission in the range from 15 keV to 10 MeV (Fig. 17). The detection of polarization in GRB041219A with INTEGRAL has opened a new observational window providing information on the physical mechanism by which the central engine of a GRB emits the huge observed energies, a presently unresolved but crucial issue (Kalemci et al., 2007; McGlynn et al., 2007; Götz et al., 2009). This GRB has been, so far, the longest and brightest detected in the Integral FOV, and the large amount of photons detected during its short duration has allowed for the first time to search with high sensitivity for polarization in the GRB soft-gamma ray emission. In fact, the INTEGRAL data have revealed a high polarization level (68 %) during the brightest part of GRB041219A (Mc Glynn et al., 2007).

Polarization detected by INTEGRAL in the high energy photons produced close to the Crab Pulsar (as described in a previous section) has been used to derive an upper limit proving the absence of vacuum birefringence effects in photon propagation, in turn constraining the Lorentz Invariance Violation (LIV) in Quantum Electro Dynamics (Maccione et al., 2008). This unprecedented limit was correlated to the parameter $|\xi| < 9 \times 10^{-10}$, derived from the INTEGRAL data with an accuracy 3 orders of magnitude better than before.

This upper limit can be constrained even further, if a source of polarized emission, more distant than the Crab, is being used: Recently, Götz et al.,(2011) determined the distance to GRB041219A, corresponding to a redshift z = 0.31 (+0.54, -0.26) using the CFHT/WirCAM instrument. The energy dependent analysis of the polarization level for this GRB has allowed to lower the existing limit by several orders of magnitude, now with a much stronger constraint for the $|\xi|$ parameter to a level of $< 1.1 \times 10^{-14}$, at least $10^5$ times better than any other limit obtained to date (Laurent et al., 2011a). The constraint on LIV violation could be improved further with similar polarization data obtained from distant and bright AGNs.

The above mentioned measurements have opened a new discovery window for the INTEGRAL observatory, now proven to be effective in the 'polarimetry window' and being able to impact on the 'new physics' science by setting important constraints on the physics describing the propagation of photons in the Universe and on the QED theory.

INTEGRAL provides valuable synergy with other current astronomical facilities, and it may be a pathfinder for future missions. New astronomical detectors dedicated to the violent Universe are going to be operational during the next years, and, will start the epoch of multi-messenger astronomy. Among others, these include large radio telescopes (LOFAR, ALMA, SKA), large Cherenkov telescope arrays (CTA), new ultra high-energy cosmic-ray detectors (AUGER), large neutrino detectors (IceCube, KM3Net), and advanced gravitational radiation detectors (Advanced LIGO, Virgo). In allowing simultaneous observations of various violent phenomena seen in other bands of the electro-magnetic spectrum or even with gravitational radiation or neutrinos, INTEGRAL has been opening new windows on the Universe.

Acknowledgement[5]

This paper is based on the scientific case made by the INTEGRAL Users Group (IUG), to support the ESA 2010 request for the extension of the INTEGRAL mission until 31 December 2014.

During 2010, members of the IUG were: A. Bazzano (INAF/IASF, Rome), T. Belloni (INAF, Brera Observ.), S. Brandt (DNSC, Copenhagen), E. Churazov (IKI, Moscow & MPA, Garching), R. Diehl (MPE, Garching), M. Falanga (ISSI, Bern), N. Gehrels (NASA/GSFC), A. Goldwurm (SAp-CEA, Gif-sur-Yvette), S. Grebenev (IKI Moscow), W. Hermsen (SRON, Utrecht), M. Hernanz (IEEC-CSIC, Barcelona), E. van den Heuvel (U. Amsterdam), P. Kretschmar (ESA/ESAC), F. Lebrun (SAp-CEA, Gif-sur-Yvette), M. Leising (U. Clemson), M. Mas-Hesse (INTA, Madrid), M. McConnell (U. New Hampshire), G. Palumbo (U. Bologna), J. Paul (SAp-CEA, Gif-sur-Yvette), K. Postnov (Sternberg Astr. Inst., Moscow), J.-P. Roques (CESR, Toulouse), N. Schartel (ESA/ESAC), R. Sunyaev (IKI, Moscow & MPA, Garching), P. Ubertini (INAF/IASF, Rome), R. Walter (ISDC, Versoix), J. Wilms (U. Erlangen-Nürnberg), C. Winkler (ESA/ESTEC).

---

[5] The final article is available at springerlink.com